# Learning to Determine the Quality of News Headlines

Amin Omidvar[1], Hossein Pourmodheji[1], Aijun An[1], and Gordon Edall[2]
*[1]Department of Electrical Engineering and Computer Science, York University, Canada*
*[2]The Globe and Mail, Canada*
*{omidvar, pmodheji, aan}@eecs.yorku.ca, GEdall@globeandmail.com*

Abstract: Today, most news readers read the online version of news articles rather than traditional paper-based newspapers. Also, news media publishers rely heavily on the income generated from subscriptions and website visits made by news readers. Thus, online user engagement is a very important issue for online newspapers. Much effort has been spent on writing interesting headlines to catch the attention of online users. On the other hand, headlines should not be misleading (e.g., clickbaits); otherwise readers would be disappointed when reading the content. In this paper, we propose four indicators to determine the quality of published news headlines based on their click count and dwell time, which are obtained by website log analysis. Then, we use soft target distribution of the calculated quality indicators to train our proposed deep learning model which can predict the quality of unpublished news headlines. The proposed model not only processes the latent features of both headline and body of the article to predict its headline quality but also considers the semantic relation between headline and body as well. To evaluate our model, we use a real dataset from a major Canadian newspaper. Results show our proposed model outperforms other state-of-the-art NLP models.

# 1 INTRODUCTION

People's attitude toward reading newspaper articles is changing in a way that people are more willing to read online news articles than paper-based ones. In the past, people bought a newspaper, saw almost all the pages while scanning headlines, and read through articles which seemed interesting (Kuiken, Schuth, Spitters, & Marx, 2017). The role of headlines was to help readers have a clear understanding of the topics of the article.

But today, online news publishers are changing the role of headlines in a way that headlines are the most important way to gain readers' attention. One important reason is that online news media publishers rely on the incomes generated from the subscriptions and clicks made by their readers (Reis et al., 2015). Furthermore, the publishers need to attract more readers than their competitors if they want to succeed in this competitive industry. The aforementioned reasons are the most important ones why some of the online news media come up with likable headlines to lure the readers into clicking on their headlines. These likable headlines may increase the number of clicks but at the same time will disappoint the readers since they exaggerate the content of the news articles (Omidvar, Jiang, & An, 2018).

Therefore, having a tool that can predict the quality of news headlines before publication would help authors to choose those headlines that not only increase readers' attention but also satisfy their expectations. However, there are some challenges to predict the quality of headlines. First, there is no labelled data set specifying the quality of headlines. Thus, given a set of articles and users' browsing history on the articles, how to determine the quality of headlines is an open issue. Second, given labelled data, how to build a model that can accurately predict the quality of headlines considering the metrics that data is labelled.

The main contributions of this research are as follows:

First, we proposed a novel headline quality detection approach for published headlines using dwell time and click count of the articles and we provide four headline quality indicators. By using this approach, we can label news article datasets of any size automatically, which is not possible by employing human annotators. Using human annotators to label data is costly, requires much time and effort, and may result in inconsistent labels/evaluation due to subjectivity. To the best of our knowledge, none of the previous related research have conducted similar approach for headline quality detection.

Second, we develop a deep network based predictive model that incorporates some advanced features of DNN to predict the quality of unpublished headlines using the previous approach as a ground truth. The proposed model considers the proposed headline quality indicators by considering the similarity between the headline and its article, and their latent features.

The rest of this paper is organized as follows. In section 2, the most relevant works regarding headline quality in the field of Computer Science and Psychology are studied. In section 3, we propose four quality indicators to represent the quality of headlines. Also, we label our dataset using a novel way to calculate the proposed quality indicators for published news articles. Next, we propose our novel deep learning architecture in section 4 to predict the headline quality for unpublished news articles. We use the calculated headline quality from the section 3 as ground truth to train our model. Then in section 5, our proposed model is compared with baseline models. Finally, this study is wrapped up with a conclusion in section 6.

# 2 RELATED WORKS

Many studies in different areas such as computer science, psychology, anthropology, and communication have been conducted on the popularity and accuracy of the news headlines over the past few years. In this section, the most relevant works in the domain of computer science and psychology are briefly described.

Researchers manually examined 151 news articles from four online sections of the El Pais, which is a Spanish Newspaper, in order to find out features which are important to catch the readers' attention. They also analysed how important linguistic techniques such as vocabulary and words, direct appeal to the reader, informal language, and simple structures are in order to gain the attention of readers (Palau-Sampio, 2016).

In another research, 2 million Facebook posts by over 150 U.S. based media organizations were examined to detect clickbait headlines. They found out clickbaits are more prevalent in entertaining categories (Rony, Hassan, & Yousuf, 2017). In order to determine the organic reach (i.e., which is the number of visitors without paid distribution) of the tweets, social sharing patterns were analysed in (Chakraborty, Sarkar, Mrigen, & Ganguly, 2017). They showed how the differences between customer demographics, follower graph structure, and type of text content can influence the tweets quality.

Ecker et al. (Ecker, Lewandowsky, Chang, & Pillai, 2014) studied how misinformation in news headlines could affect news readers. They found out headlines have an important role to shape readers' attitudes toward the content of news. In (Reis et al., 2015), they extracted features from the content of 69907 news articles in order to find approaches which can help to attract clicks. They discovered the sentiment of the headline is strongly correlated to the popularity of the news article.

Some distinctive characteristics between accurate and clickbait headlines in terms of words, entities, sentence patterns, paragraph structures etc. are discovered in (Chakraborty, Paranjape, Kakarla, & Ganguly, 2016). At the end, they proposed an interesting set of 14 features to recognize how accurate headlines are. In another work, linguistically-infused network was proposed to distinguish clickbaits from accurate headlines using the passages of both article and headline along with the article's images (Glenski, Ayton, Arendt, & Volkova, 2017). To do that, they employed Long Short-Term Memory (LSTM) and Convolutional Neural Network architectures to process text and image data, respectively.

One interesting research measured click-value of individual words of headlines. Then they proposed headline click-based topic model (HCTM) based on latent Dirichlet allocation (LDA) to identify words that can bring more clicks for headlines (J. H. Kim, Mantrach, Jaimes, & Oh, 2016). In another related research (Szymanski, Orellana-Rodriguez, & Keane, 2017), a useful software tool was developed to help authors to compose effective headlines for their articles. The software uses state of the art NLP techniques to recommend keywords to authors for inclusion in articles' headline in order to make headlines look more intersecting. They calculated two local and global popularity measures for each keyword and use supervised regression model to predict how likely headlines will be widely shared on social media.

Deep Neural Networks has become a widely used technique that has produced very promising results in news headline popularity task in recent years (Bielski & Trzcinski, 2018; Stokowiec, Trzciński, Wołk, Marasek, & Rokita, 2017; Voronov, Shen, & Mondal, 2019). Most NLP approaches employ deep learning models and they do not usually need heavy feature engineering and data cleaning. However, most of the traditional methods rely on the graph data of the interactions between users and contents.

For detecting clickbait headlines, lots of research have been conducted so far (Fu, Liang, Zhou, & Zheng, 2017; Venneti & Alam, 2018; Wei & Wan, 2017; Zhou, 2017). In (Martin Potthast, Tim Gollub, Matthias Hagen, 2017) they launched a clickbait challenge competition and also released two supervised and unsupervised datasets which contains over 80000 and 20000 samples, respectively. Each sample contains news content such as headline, article, media, keywords, etc. For the supervised dataset, there are five scores from five different judges in a scale of 0 to 1. A leading proposed model in the clickbait challenge competition (Omidvar et al., 2018), which its name is albacore in the published result list on the competition's website[1], employed bi-directional GRU along with Fully connected NN layers to determine how much clickbait each headline is. They showed that posted headline on the twitter (i.e., postText field) is the most important features of each sample to predict the judges' score due to the fact that maybe human evaluators only used posted Headline feature to label each sample. The leading approach not only got the first rank in terms of Mean Squared Error (MSE) but also is the fastest among all the other proposed models.

To the best of our knowledge, none of the previous studies analysed the quality of headlines by considering both their popularity and truthfulness (i.e., non clickbait). The reason is that almost all of the previous research, especially those for clickbait detection, looked at the problem as a binary classification task. Also, most of them depend on human evaluators to label the dataset. In our proposed data labelling approach, we determine the quality of headlines based on 4 quality indicators by considering both their popularity and validity. Also, we come up with a novel approach to calculate 4 quality indicators automatically by using users' activity log dataset. Then, our trained deep learning model not only determines how popular headlines are, but also how honest and accurate they are.

# 3 LABELING DATA

In this section, a novel approach is introduced to calculate the quality of published headlines based on users' interactions with articles. This approach is used for labeling our dataset.

## 3.1 Data

Our data is provided by The Globe and Mail which is a major Canadian newspaper. It contains a news corpus dataset (containing articles and their metadata) and a log dataset (containing interactions of readers with the news website). Every time a reader opens an article, writes a comment or takes any other trackable action, it is detected on the website, and then is stored as a record in a log data warehouse. Generally, every record contains 246 captured attributes such as event ID, user, time, date, browser, IP address, etc.

The log data can give useful insights into readers' behaviours. However, there are noise and inconsistencies in the clickstream data which should be cleaned before calculating any measures, applying any models, or extracting any patterns. For example, users may leave articles open in the browser for a long time while doing other activities, such as browsing other websites in another tab. In this case, some news articles will get high fake dwell times from some readers.

There are approximately 2 billion records of users' actions in the log dataset. We use the log dataset to find how many times each article has been read and how much time users spent reading it. We call these two measures click count and dwell time, respectively.

## 3.2 Quality Indicators

Due to the high cost of labelling supervised training data using human annotators, large datasets are not available for most NLP tasks (Cer et al., 2018).

In this section, we calculate the quality of published articles using articles' click count and dwell time measures. By using the proposed approach, we can label any size of database automatically and use those labels as ground truths to train deep learning models. A dwell time for article *a* is computed using Formula 1.

$$D_a = \frac{\sum_u T_{a,u}}{C_a} \quad (1)$$

where $C_a$ is the number of times article *a* was read and $T_{a,u}$ is the total amount of time that user *u* has spent reading article *a*. Thus, the dwell time of article *a* (i.e., $D_a$) is the average amount of time spent on the article during a user visit. The values of read count and dwell time are normalized in the scale of zero to one.

[1] https://www.clickbait-challenge.org/#results

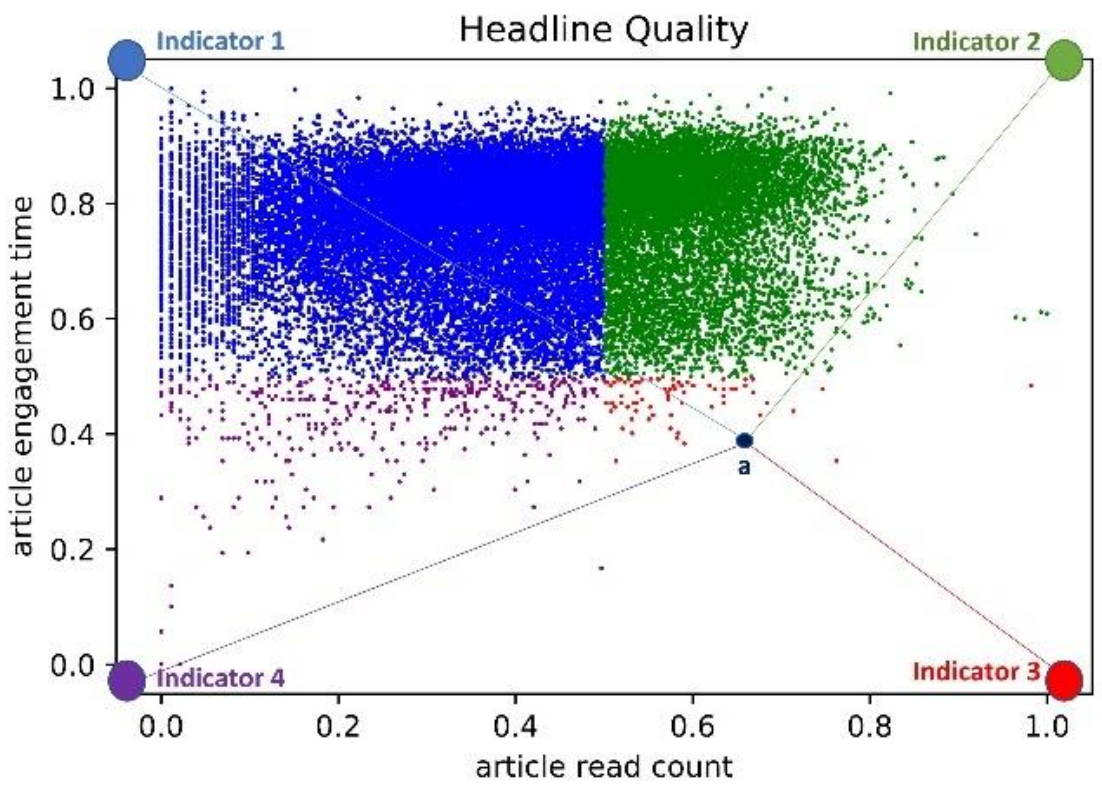

Figure 1: Representing News Headlines' quality with respect to the four quality indicators.

By considering these two measures for headline quality, we can define four quality indicators which are shown by the 4 corners of the rectangle in Figure 1. We did not normalize articles' dwell time by articles' length since the correlation and mutual information between articles' reading time and articles' length were 0.2 and 0.06, respectively which indicates there is a very low dependency between these two variables in our dataset.

- **Indicator 1:** High dwell time but low read count. Articles close to this indicator were interesting for users because of their high dwell time but their headlines were not interesting enough to motivate users to click on the articles. However, those users who read these articles spent a significant amount of time reading them.
- **Indicator 2:** High dwell time and high read count. Articles close to indicator 2 had interesting headlines since they had opened by many users, and the articles were interesting as well because of their high dwell time.
- **Indicator 3:** Low dwell time but high read count. Articles close to this indicator have high read count but low dwell time. These headlines were interesting for users, but their articles were not. We call this type of headlines misleading headlines since the articles do not meet the expectation of the readers. As we can see in Figure 1, very few articles reside in this group.
- **Indicator 4:** Low dwell time and read count. Headlines of these articles were not successful to attract users and those who read them did not spend much time reading them.

The probability that article *a* belongs to each quality indicator *i* (i.e. $P_{a,i}$) is calculated using formula 2 which $\| \|_2$ is the $L_2$ norm. Softmax function is used to convert the calculated similarities into probabilities.

$$\begin{bmatrix} P_{a,1} \\ P_{a,2} \\ P_{a,3} \\ P_{a,4} \end{bmatrix} = Softmax \begin{pmatrix} \sqrt{2} - \|(C_a, (1 - D_a))\|_2 \\ \sqrt{2} - \|((1 - C_a), (1 - D_a))\|_2 \\ \sqrt{2} - \|((1 - C_a), D_a)\|_2 \\ \sqrt{2} - \|(C_a, D_a)\|_2 \end{pmatrix} \quad (2)$$

# 4 PREDICT HEADLINE QUALITY

In this section, we propose a novel model to predict the quality of unpublished news headlines. To the best of our knowledge, we are the first to consider latent features of headlines, bodies, and the semantic relation between them to find the quality of news headlines.

## 4.1 Problem Definition

We consider the task of headline quality prediction as a multiclass classification problem. We assume our input contains a dataset $D = \{(H_i, A_i)\}_i^N$ of $N$ news articles that each news article contains a header and an article which are shown by $H_i$ and $A_i$, respectively. An approach for learning the quality of headline is to define a conditional probability $P(I_j \mid H_i, A_i, \theta)$ for each quality indicator $I_j$ with respect to the header text (i.e., $H_i = \{t_1, t_2, \ldots, t_K\}$), article text (i.e., $A_i = \{z_1, z_2, \ldots, z_m\}$), and parameterized by a model with parameters $\theta$. We then estimate our prediction for each news article in our database as:

$$\hat{y}_i = argmax_{\, j \in \{1,2,3,4\}} P(C_j \mid H_i, A_i, \theta) \quad (3)$$

## 4.2 Proposed Model

In this section we propose a deep learning model to predict the quality of headlines before publication. The proposed model is implemented in python language and will be put on authors' GitHub account after paper publication. The architecture of the proposed model is illustrated in Figure 2.

### 4.2.1 Embedding Layer

This layer, which is available in Keras library[2], converts the one-hot-encoding of each word in headlines and articles to the dense word embedding vectors. The embedding vectors are initialized using GloVe Embedding vectors (Pennington, Socher, & Manning, 2014). We find that 100-dimensional embedding vectors lead to the best result. Also, we use a drop out layer on top of the embedding layer to drop 0.2 percent of the output units.

### 4.2.2 Similarity Matrix Layer

Because one of the main characteristics of high-quality headlines is that a headline should be related to the body of its article, the main goal of this layer is to find out how related each headline is to the article's body. Embedding vectors of the words of both the headline and the first paragraph of the articles are the inputs to this layer. We use the first paragraph of the article since the first paragraph is used extensively for news summarizing task due to its high importance to representing the whole news article (Lopyrev, 2015).

In Figure 2, each cell $c_{i,j}$ represents the similarity between words $h_i$ and $b_j$ from the headline and its article, respectively, which is calculated using the cosine similarity between their embedding vectors using formula 4.

$$C_{ij} = \frac{\vec{z}_i^{\,T}\,\vec{t}_j}{\|\vec{z}_i\|.\,\|\vec{t}_j\|} \tag{4}$$

Using the cosine similarity function will enable our model to capture the semantic relation between the embedding vectors of two words $z_i$ and $t_j$ in the article and header, respectively. Also, the 2-d similarity matrix allows us to use 2-d CNN which has shown great performance for text classification through abstracting visual patterns from text data (Pang et al., 2016). In fact, matching headline and article is viewed as image recognition problem and 2-d CNN is used to solve it.

### 4.2.3 Convolution and Max-Pooling Layers

Three Convolutional Network layers, each of which contains 2-d CNN and 2-d Max-Pooling layers, are used on top of the similarity matrix layer. The whole Similarity Matrix is scanned by the first layer of 2-d CNN to generate the first feature map. Different level of matching patterns is extracted from the Similarity Matrix in each Convolutional Network Layer based on the formula 5.

$$x_{i,j}^{(l+1,p)} = f\left(\sum_{p=1}^{n_l}\left(\sum_{y=1}^{v_k}\left(\sum_{m=1}^{v_k} w_{y,m}^{(l+1,k)}.\ x_{i+y,j+m}^{(l,p)}\right) + b^{(l+1,p)}\right)\right) \tag{5}$$

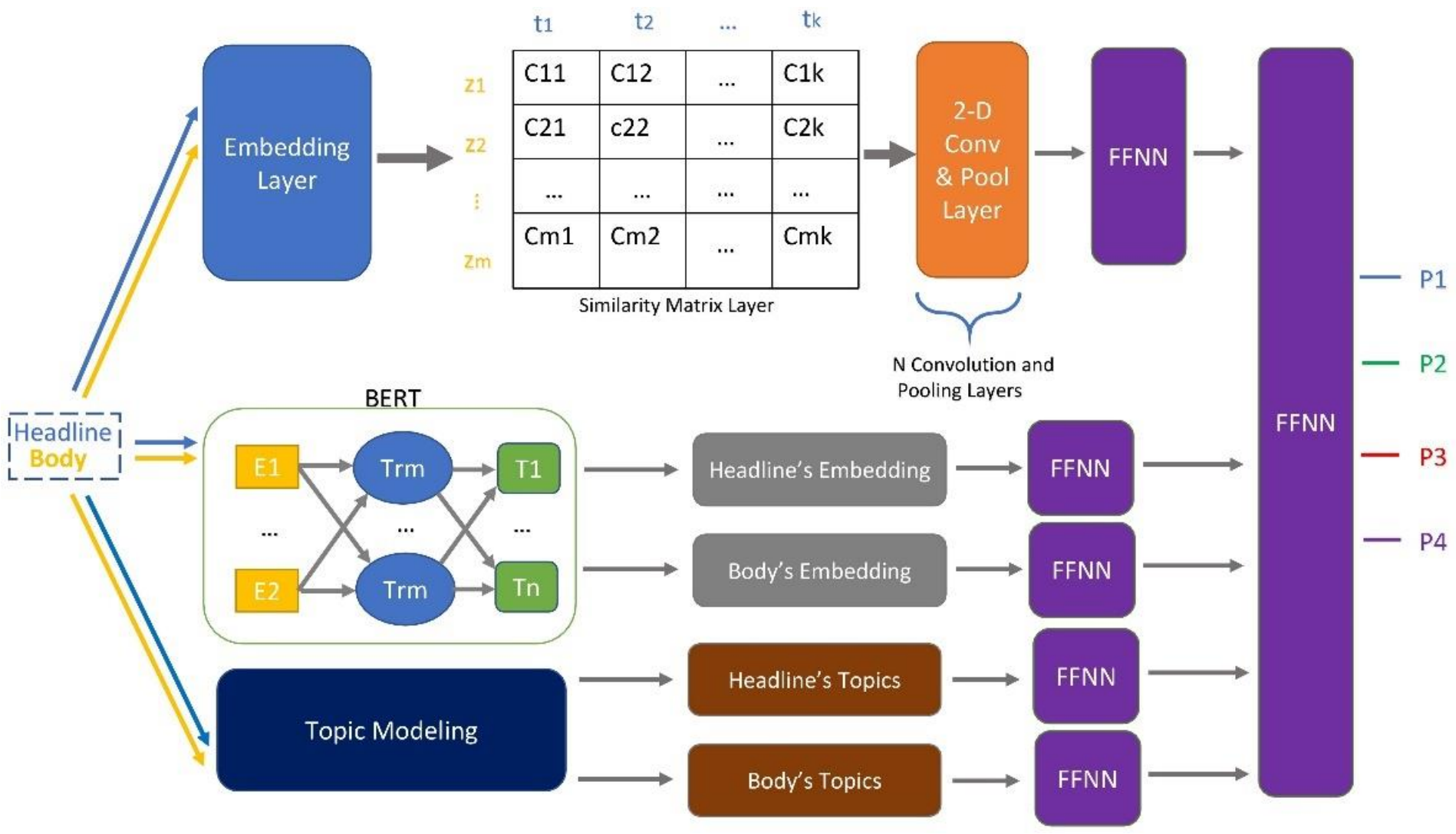

Figure 2: The proposed model for predicting news headlines' quality according to the four quality indicators.

[2] https://keras.io/layers/embeddings/

$x^{(l+1)}$ is the computed feature map at level $l+1$, $w^{(l+1,k)}$ is the $k$-th square kernel at the level $l+1$ which scans the whole feature map $x^{(l)}$ from the previous layer, $v_k$ is the size of the kernel, $b^{(l+1)}$ are the bias parameters at level $l+1$, and ReLU (Dahl, Sainath, & Hinton, 2013) is chosen to be the activation function *f*. Then we will get feature maps by applying dynamic pooling method (Socher, Huang, Pennington, Ng, & Manning, 2011).

We use (5*5), (3*3), and (3*3) for the size of kernels, 8, 16, and 32 for the number of filters, and (2*2) for the pool size in each Convolutional Network layer, respectively. The result of the final 2-d Max-Pooling layer is flattened to the 1-d vector. Then it passes a drop out layer with the rate of 0.2. In the end, the size of the output vector is reduced to 100 using a fully-connected layer.

#### 4.2.4 BERT

Google's Bidirectional Encoder Representations from Transformers (BERT) (Devlin, Chang, Lee, & Toutanova, 2018) is employed to transform variable-length inputs, which are headlines and articles, into fixed-length vectors for the purpose of finding the latent features of both headlines and articles. BERT's goal is to produce a language model using the Transformer model. Details regarding how Google Transformer works is provided in (Vaswani et al., 2017).

BERT is pre-trained on a huge dataset to learn the general knowledge that can be used and combined with the acquire knowledge on a small dataset. We use the publicly available pre-trained BERT model (i.e., BERT-Base, Uncased)[3], published by Google. After encoding each headline into a fixed-length vector using BERT, a multi-layer perceptron is used to project each encoded headline into a 100-d vector. The same procedure is performed for the articles as well.

#### 4.2.5 Topic Modelling

We use None Negative Matrix Factorization (NNMF) (J. Kim, He, & Park, 2014) and Latent Dirichlet Allocation (LDA) (Hoffman, Blei, & Bach, 2010) from Scikit-learn library to find topics from both headlines and articles. Since headlines are significantly shorter than articles, we use separate topic models for headlines and articles. Even though both NNMF and LDA can be used for topic modelling their approach is totally different from each other in a way that the former is based on linear algebra and the latter relies on probabilistic graphical modelling. We find out NNMF extracts more meaningful topics than LDA on our news dataset.

We create matrix $A$, in which each article is represented as a row and columns are the TF-IDF values of article's words. TF-IDF is an acronym for term frequency - inverse document frequency which is a statistical measure to show how important a word is to an article in a group of articles. Term Frequency (TF) part calculates how frequently a word appears in an article divided by the total number of words in that article. The Inverse Document Frequency (IDF) part weighs down the frequent words while scaling up the rare words in an entire corpus.

Then we use NNMF to factorize matrix $A$ into two matrices $W$ and $H$ which are document to topic matrix and topic to word matrix, respectively. When these two matrices multiplied, the result is the matrix $A$ with the lowest error (formula 6).

$$A_{n \times v} = W_{n \times t}\, H_{t \times v} \tag{6}$$

In formula 6, $n$ is the number of articles, $v$ is the size of vocabulary, and $t$ is the number of topics ($t \ll v$) which we set it to 50. As it is shown in Figure 2, we use topics (i.e., each rows of matrix $W$) as input features to the Feedforward Neural Network (FFNN) part of our model.

#### 4.2.6 FFNN

As we can see in Figure 2, FFNN layers are used in different parts of our proposed model. The rectifier is used as the activation function of all layers except the last one. The activation function of the last layer is softmax which calculates the probability of the input example being in each quality indicator. We find that using a batch normalization layer before the activation layer in all layers helps to reduce the loss of our model since a batch normalization layer normalizes the input to the activation function so that the data are centred in the linear part of the activation function.

# 5 RESULTS

## 5.1 Baselines

For evaluation, we have compared our proposed model with the following baseline models.

[3] https://github.com/google-research/bert\#pre-trained models

### 5.1.1 EMB + 1-d CNN + FFNN

This embedding layer is similar to the embedding layer of the proposed model which will convert one-hot representation of the words to the dense 100-d vectors. A drop out layer is used on top of the embedding layer to drop 0.2 percent of the output units. Also, we use GloVe embedding vectors to initialize word embedding vectors (Pennington et al., 2014). The next layer is 1-d CNN which works well for identifying patterns within single spatial dimension data such as text, time series, and signal. Many recent NLP models employed 1-d CNN for text classification tasks (Yin, Kann, Yu, & Schütze, 2017). The architecture is comprised of two layers of convolution on top of the embedding layer. The last layer is a single layer FFNN using softmax as its activation function.

### 5.1.2 Doc2Vec + FFNN

Doc2Vec[4] is an implementation of the Paragraph Vector model, which was proposed in (Le & Mikolov, 2014). It is an unsupervised learning algorithm that can learn fixed-length vector representations for different length pieces of text such as paragraphs and documents. The goal is to learn the paragraph vectors by predicting the surrounding words in contexts obtained from the paragraph. It consists of two different models which are Paragraph Vector Distributed Memory Model (PV-DMM) and Paragraph Vector without word ordering Distributed bag of words (PV-DBOW). The former has much higher accuracy than the latter but the combination of them yields to the best result.

We convert headlines and bodies into two separate 100-d embedded vectors. These vectors are fed into FFNN, which comprises of two hidden layers with the size of 200 and 50 consecutively. ReLU is used for the activation function of all FFNN layers except the last layer which employs softmax function.

### 5.1.3 EMB + BGRU + FFNN

This is a Bidirectional Gated Recurrent Unit on top of the Embedding layer. GRU employs two gates to trail the input sequences without using separate memory cells which are reset $r_t$ and update $z_t$ gates, respectively.

$$r_t = \sigma(W_r x_t + U_r h_{t-1} + b_r) \quad (7)$$

$$z_t = \sigma(W_z x_t + U_z h_{t-1} + b_z) \quad (8)$$

$$h_t^{\sim} = tanh(W_h x_t + r_t * (U_h h_{t-1}) + b_h) \quad (9)$$

$$h_t = (1 - z_t) * h_{t-1} + z_t * h_t^{\sim} \quad (10)$$

In formulas 7 and 8, $W_r$, $U_r$, $b_r$, $W_z$, $U_z$, $b_z$ are the parameters of GRU that should be trained during the training phase. Then, the candidate and new states will be calculated at time $t$ based on the formula 9 and 10, respectively.

In formulas 9 and 10, * denotes an elementwise multiplication between the reset gate and the past state. So, it determines which part of the previous state should be forgotten. And update gate in formula 10 determines which information from the past should be kept and which one should be updated. The forward way reads the post text from $x_1$ to $x_N$ and the backward way reads the post text from $x_N$ to $x_1$.

$$\overrightarrow{h_n} = \overrightarrow{GRU}(x_n, \overrightarrow{h_{n-1}}) \quad (11)$$

$$\overleftarrow{h_n} = \overleftarrow{GRU}(x_n, \overleftarrow{h_{n+1}}) \quad (12)$$

$$h_n = [\overrightarrow{h_n}, \overleftarrow{h_n}] \quad (13)$$

And the input to the FFNN layer is the concatenation of the last output of forward way and backward way.

### 5.1.4 EMB + BLSTM + FFNN

This is a Bidirectional Long Short Term Memory (LSTM) (Hochreiter & Schmidhuber, 1997) on top of the Embedding layer. Embedding and FFNN layers are similar to the previous baseline. The only difference here is using LSTM instead of using GRU.

## 5.2 Evaluation Metrics

Mean Absolute Error (MAE) and Relative Absolute Error (RAE) are used to compare the result of the proposed model with the result of the baseline models on test dataset. As we can see in formula 14, RAE is relative to a simple predictor, which is just the average of the ground truth. The ground truth, the predicted values and the average of the ground truth are shown by $P_{ij}$, $\hat{P}_{ij}$, and $\bar{P}_j$, respectively.

$$RAE = \frac{\sum_{i=1}^{N}\sum_{j=1}^{4}|P_{ij} - \hat{P}_{ij}|}{\sum_{i=1}^{N}\sum_{j=1}^{4}|P_{ij} - \bar{P}_j|} \quad (14)$$

## 5.3 Experimental Results:

We train our proposed model with the same configuration two times: once using hard labels (i.e., assigning a label 1 and three 0s to the quality indicators for each sample) and the other time using soft labels, which were calculate by formula 2. We use categorical cross entropy loss function for the former and MSE loss function for the latter. Then we find out our proposed model will be trained more efficiently by using soft targets than using hard targets, same as what was shown in (Hinton, Vinyals, & Dean, 2015). The reason could be that soft targets provide more information per training example in comparison with hard targets and much less variance in the gradient between our training examples. For instance, for the machine learning tasks such as MNIST (LeCun, Bottou, Bengio, & Haffner, 1998), one case of image 1 may be given probabilities $10^{-6}$ and $10^{-9}$ for being 7 and 9, respectively while for another case of image 1 it may be the other way round. So, we decide to train our proposed model and all the baseline models just using soft targets.

[4] https://radimrehurek.com/gensim/models/doc2vec.html

The results of the proposed model and baseline models on the test data are shown in Table 1. The loss function of the proposed model and all the baseline models is based on the MSE between the predicted quality indicators by the models and the ground truth calculated in section 3.2 (Soft labels). And we use Adam optimizer for the proposed model and all our baseline models (Kingma & Ba, 2015). Also, we split our dataset into train, validation, and test sets using 70, 10, and 20 percent of data, respectively.

Our proposed model got the best results by having the lowest RAE among all the other models. Surprisingly, TFIDF performs better than the other baseline models. It can be due to the fact that the number of articles in our dataset is not big (28751), so complex baseline models may overfit to the training data set.

Also, we are interested in finding the importance of the latent features regarding the semantic relation between headlines and articles. So, we have removed embedding, similarity matrix, and 2-D CNN layers from the proposed model. After making these changes, RAE was increased by 6 percent in comparison with the original proposed model. This shows that measuring the similarity between article's headline and body is beneficial for headline quality prediction.

Table 1: Comparison between the proposed model and baseline models.

| Models | MAE | RAE |
|---|---|---|
| EMB + 1-D CNN + FFNN | 0.044 | 105.08 |
| Doc2Vec + FFNN | 0.043 | 101.61 |
| EMB + BLSTM + FFNN | 0.041 | 97.92 |
| EMB + BGRU + FFNN | 0.039 | 94.38 |
| TF-IDF + FFNN | 0.038 | 89.28 |
| Proposed Model without Similarity Matrix | 0.036 | 86.1 |
| **Proposed Model** | **0.034** | **80.56** |

# 6 CONCLUSION

In this research, we proposed a method for calculating the quality of the published news headlines with regard to the four proposed quality indicators. Moreover, we proposed a novel model to predict the quality of headlines before their publication, using the latent features of headlines, articles, and their similarities. The experiment was conducted on a real dataset obtained from a major Canadian newspaper. The results showed the proposed model outperformed all the baselines in terms of Mean Absolute Error (MAE) and Relative Absolute Error (RAE) measures. As headlines play an important role in catching the attention of readers, the proposed method is of great practical value for online news media.

# ACKNOWLEDGEMENTS

This work is funded by Natural Science and Engineering Research Council of Canada (NSERC), The Globe and Mail, and the Big Data Research, Analytics and Information Network (BRAIN) Alliance established by the Ontario Research Fund – Research Excellence Program (ORF-RE).